\documentclass{ws-ijmpd}
\usepackage[dvips]{color}
\usepackage{graphicx}
\usepackage{dcolumn}
\usepackage{bm}

\newcommand\ignore[1]\relax

\newcommand{\tr}{{\rm tr}}

\newcommand{\mev}{{\rm MeV}}
\newcommand{\gev}{{\rm GeV}}
\newcommand{\tev}{{\rm TeV}}

\newcommand{\lmd}{\lambda}
\newcommand{\TeV}{\mbox{TeV}}
\newcommand{\GeV}{\mbox{GeV}}

\newcommand{\tz}{{T^0}}

\newcommand{\tzr}{{T_r^0}}
\newcommand{\tzi}{{T_i^0}}

\begin{document}

\markboth{T.~Araki,\,C.Q.~Geng,\,and K.I.~Nagao}
{Signature of Dark Matter in Inert Triplet Higgs Model}

%
%

\title{{Signatures of Dark Matter in Inert Triplet Models}}

\author{Takeshi~Araki$^1$,
C.~Q.~Geng$^{2,3}$
and Keiko~I.~Nagao$^2$\footnote{Talk presented by K.I. Nagao at the
2nd International Workshop on Dark Matter, Dark Energy and Matter-Antimatter Asymmetry, Hsinchu, Taiwan, 5-6 Nov 2010.
}}
\address{$^1$Institute of High Energy Physics, Chinese Academy of Sciences, Beijing 100049, China\\
$^2$Department of Physics, National Tsing Hua University, Hsinchu, Taiwan 300\\
$^{3}$National Center for Theoretical Sciences, Hsinchu, Taiwan 300}

\maketitle


\begin{abstract}
In this talk, we will review the signatures of dark matter in two inert triplet models. For the first model with the hypercharge Y=0, 
the dark matter mass around $5.5~\TeV$ is favored by both the WMAP data and the direct detection. In contrast, for the second model of Y=2, 
 it is excluded by the direct detection experiments although dark matter with its mass around $2.8~\TeV$ is allowed by WMAP.
\end{abstract}

\keywords{dark matter; inert triplet model.}

\section{Introduction}
Since the standard model (SM) cannot give an explanation for
dark matter, new physics is expected.
In this talk, we review the signatures of dark matter in  two models  which contain $SU(2)_L$ triplet scalars with the hypercharges Y=0 and 2
under $U(1)_Y$, respectively \cite{ITM}.
In these models, the triplets are odd under an $Z_2$ symmetry so that they neither directly couple to the SM fermions nor develop 
vacuum expectation values (VEVs), while the neutral components of the triplets are the dark particles.
We will refer to the models as the inert triplet models (ITMs).
%
%
The number of new parameters in the Y=0 ITM is three compared to the SM, which is the same as those in the inert singlet model  \cite{Silveira:1985rk}. 
Clearly, the Y=0 ITM is one of the minimal inert models. 
Similarly, the Y=2 ITM like 
the inert doublet Model~\cite{Ma:2006km}$^{-}$\cite{Hambye:2009pw} has five new parameters.

The relic abundance of the cold dark matter in the universe is determined to be~\cite{Komatsu:2010fb}
\begin{eqnarray}
\Omega_{CDM} h^2=0.1123 \pm 0.0035,
\label{eq:WMAP_relic}
\end{eqnarray}
where $h = 0.710 \pm 0.025$ is the scaled current Hubble parameter in units of $100\,\mathrm{km\, sec^{-1}\, Mpc^{-1}}$.
On the other hand, the direct searches  also provide constraints on dark matter.
For example, 
the spin-independent (SI) dark matter cross section has to satisfy~\cite{Angle:2007uj}
\begin{eqnarray}
\sigma_{SI}\lesssim (5\times 10^{-44} )- 10^{-42}\,.
\label{eq:DD_constraint}
\end{eqnarray}
Note that to have the above constraint,  we have assumed that dark matter is the Weakly Interacting Massive Particle (WIMP) 
with its mass smaller than $O(10^3)~\GeV$~\cite{Angle:2007uj}.
%


\section{Inert Triplet Models}
\label{sec:models}
\subsection{Inert Triplet Model with Y=0}
\label{subsec:ITMY0}
We extend the SM by adding a zero hypercharge $SU(2)_L$ 
triplet scalar with an unbroken $Z_2$ symmetry. 
The relevant Lagrangian
is given by
\begin{eqnarray}
{\cal L}&=&|D_\mu H|^2 + \tr|D_\mu T|^2 - V(H,T), 
\nonumber\\
V(H,T) &=& m^2 H^\dag H + M^2 \tr[T^2] + \lambda_1 |H^\dag H|^2
+ \lambda_2 \left(\tr[T^2]\right)^2 + \lambda_3 H^\dag H\ \tr[T^2]\,, 
\label{eq:3vertex}
\end{eqnarray}
where $D_\mu$ is the covariant derivative including the SM 
gauge bosons,
and the SM Higgs doublet $H$ and the triplet 
 $T$ scalars are defined as
\begin{eqnarray}
H=\left(\begin{array}{c} 
   \phi^+ \\ 
   \frac{1}{\sqrt{2}}(h +i\eta)
  \end{array}\right),\ \ \ 
T=
\left(\begin{array}{cc}
   \frac{1}{\sqrt{2}}T^0 & -T^{+} \\
   -T^- & -\frac{1}{\sqrt{2}}\tz
  \end{array}\right),
\end{eqnarray}
respectively, with 
 $\langle h\rangle=v=246 ~\gev$ and 
$\langle \tz\rangle=0$. 
The stability condition of the Higgs potential requires
\begin{eqnarray}
 \lambda_1,\ \ \lambda_2 > 0,\ \
 2\sqrt{\lambda_1 \lambda_2} > |\lambda_3|\ \ {\rm for\ negative}\ \lambda_3.
\end{eqnarray}
The potential in Eq. (\ref{eq:3vertex}) becomes a local minimum if and only if
\begin{eqnarray}
m^2 <0\ ,\ \ \ 2M^2 + \lambda_3 v^2 > 0\ ,
\end{eqnarray}
where $v^2 = - m^2/\lambda_1$.
After $h$ acquires the VEV, the scalars gain 
the following masses:
\begin{eqnarray}
m_h^2 = 2\lambda_1 v^2\,,\ 
m_{\tz}^2 = m_{T^\pm}^2 = M^2 + \frac{1}{2}\lambda_3 v^2\,,
\end{eqnarray}
at tree level.
Due to the
radiative corrections \cite{Cirelli:2009uv},
the masses of $\tz$ and $T^\pm$ are split as
\begin{eqnarray}
m_{T^\pm}=m_{\tz} + (166\ \mev). \label{eq:gap}
\end{eqnarray}
Hence, $T^0$ turns out to be the lightest component of 
the triplet scalar with its stability  protected by the $Z_2$ symmetry.

Since the triplet scalar is added to the SM, one may 
expect that it affects the so-called oblique 
(S and T) parameters.
In general, however, an $Y=0$ triplet 
has no
contribution  to 
the S parameter, while the contribution to the T parameter is 
also vanishing in the limit of $m_{T^0} = m_{T^\pm}$ \cite{oblq}.
Even if we consider
the small mass splitting in 
Eq. (\ref{eq:gap}), 
its effect is negligibly small, such that
\begin{eqnarray}
T=\frac{1}{4\pi c_w^2 s_w^2 m_Z^2}
  \left[ 
        (m_\tz^2 + m^2_{T^\pm}) 
       -\frac{2m_\tz^2 m^2_{T^\pm}}{m_\tz^2 - m^2_{T^\pm}} 
        \log\frac{m^2_\tz}{m^2_{T^\pm}}
  \right] \simeq 0,
\end{eqnarray}
where $s_w(c_w)$ is the weak mixing angle and $m_Z$ 
is the $Z$ boson mass.
Therefore, the constraint on the Higgs boson mass ($m_h$) from 
the precision electroweak measurements is the 
same as that in the SM.
In our calculation, we restrict $m_h$
to be within the range of
\begin{eqnarray}
114\ \gev<m_h<185\ \gev
\end{eqnarray}
as estimated in Ref. \cite{LEP} 
with the excluded region of $158 \sim 175\ \gev$ 
reported by the Tevatron \cite{tev}.

\subsection{Inert Triplet Model with $Y=2$}
\label{subsec:ITMY2}
In the model with  the inert triplet scalar of $Y=2$,
 the $Z_2$ invariant scalar potential is 
given by
\begin{eqnarray}
V (H,T)&=& 
    m^2 H^\dag H + M^2 \tr[T^\dag T] + \lambda_1 |H^\dag H|^2 
  + \lambda_2 \tr[T^\dag T T^\dag T] + \lambda_3 \left( \tr[T^\dag T] \right)^2
  \nonumber \\
&&+ \lambda_4 H^\dag H\ \tr[T^\dag T] + \lambda_5 H^\dag T T^\dag H\ ,
\end{eqnarray}
where
\begin{eqnarray}
T=\left(\begin{array}{cc}
   \frac{1}{\sqrt{2}}T^+ & T^{++} \\
   \tzr + i\tzi & -\frac{1}{\sqrt{2}}T^+
  \end{array}\right).
\end{eqnarray}
The masses of the scalars are found to be
\begin{eqnarray}
&& m_h^2 = 2\lambda_1 v^2\ , 
\nonumber\\
&& m_\tzr^2 = m_\tzi^2 = M^2 + \frac{1}{2}(\lambda_4 + \lambda_5)v^2\ ,
\nonumber \\
&& m_{T^{\pm}}^2 = M^2 + \frac{1}{2}\left(\lambda_4 + \frac{\lambda_5}{2} \right)v^2 
                 = m_{\tzr(\tzi)}^2 - \frac{\lambda_5}{4}v^2\ ,
                 \nonumber\\
&& m_{T^{\pm\pm}}^2 = M^2 + \frac{1}{2}\lambda_4 v^2 = m_{\tzr(\tzi)}^2 - \frac{\lambda_5}{2} v^2\,.
\end{eqnarray}
%
It is clear that $\tzr$ can be the lightest $Z_2$ odd particle
if  $\lambda_5 < 0$. Since dark matter must not be a charged particle, we will concentrate on
 $\lambda_5 < 0$ afterward.

\section{Signatures of Dark Matter}
\label{sec:DM}
\subsection{Dark Matter in the model of Y=0}
\label{subsec:DM_0}
\ignore{
We now examine  the thermal relic abundance of $\tz$.
The evolution of the number density of $\tz$ is obtained by solving the 
Boltzmann equation
\begin{eqnarray}
\frac{dn_\tz}{dt}+3Hn_\tz=-\langle \sigma v_\tz\rangle (n_\tz^2-n_{\tz,\,eq}),
\label{eq:Boltzmann}
\end{eqnarray}
where $H$ is
the Hubble parameter, $v_\tz$ stands for relative
velocity of\, $T^0$, 
$\langle \cdots \rangle$ represents the thermal average of a function 
in brackets, and $n_\tz$, $n_{\tz,\,eq}$ and $\sigma$
are the number density, the number density  in thermal equilibrium
 and the total annihilation cross section of $\tz$, respectively.
}
In the model, since the masses of dark matter ($T^0$) and charged components ($T^\pm$) 
are almost degenerate,
 the coannihilation effects of $\tz\,T^\pm$ and $T^\pm \,T^\mp$ should
be included in the calculation of the relic abundance of $\tz$~\cite{Griest:1990kh}.

\begin{figure}[tbp]
\begin{center}
\includegraphics[width=10.5cm,clip]{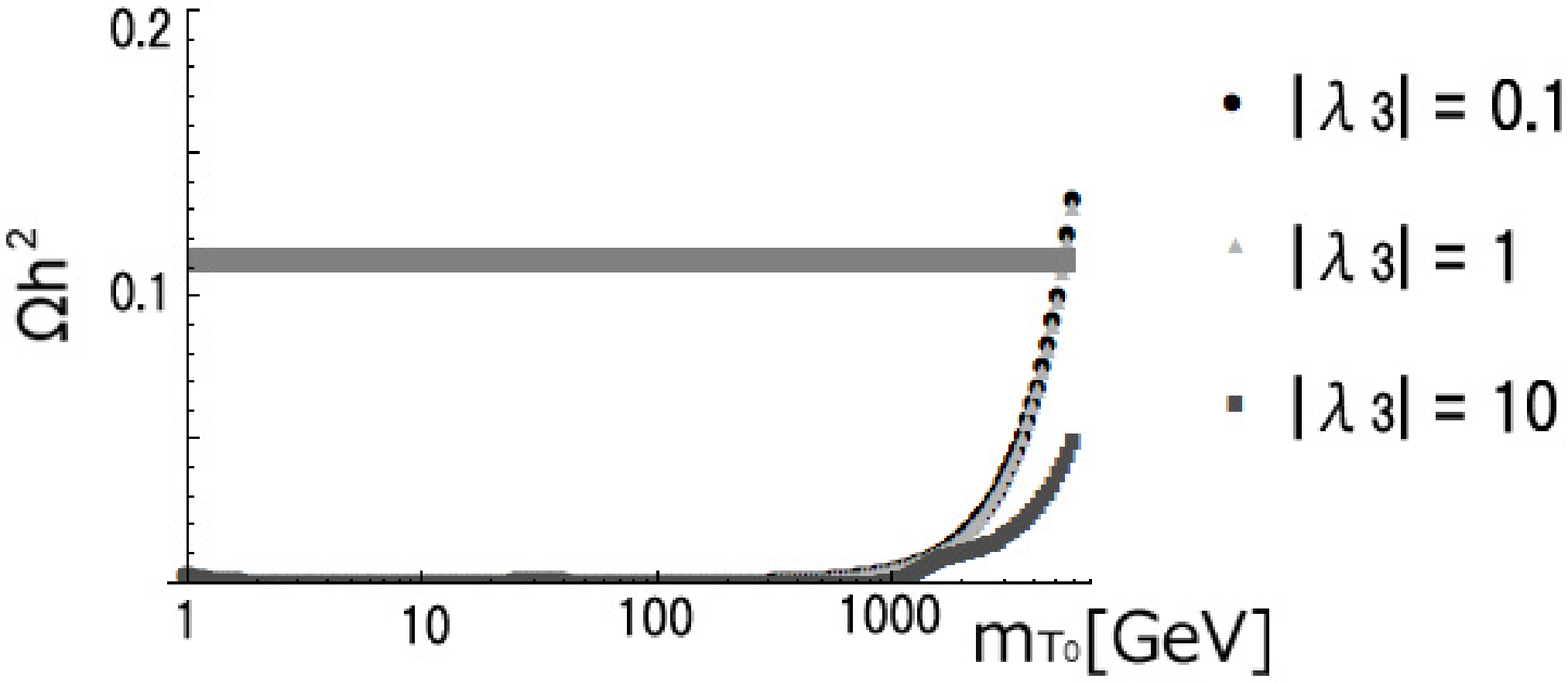}
\caption{Relic abundance for $m_h=120~\GeV$ with three different choices of $\lambda_3$, where the dark gray band represents the region 
 favored by WMAP.
 \vspace{1cm} }
\label{fig:relic_abundance}
\includegraphics[width=5.5cm,clip]{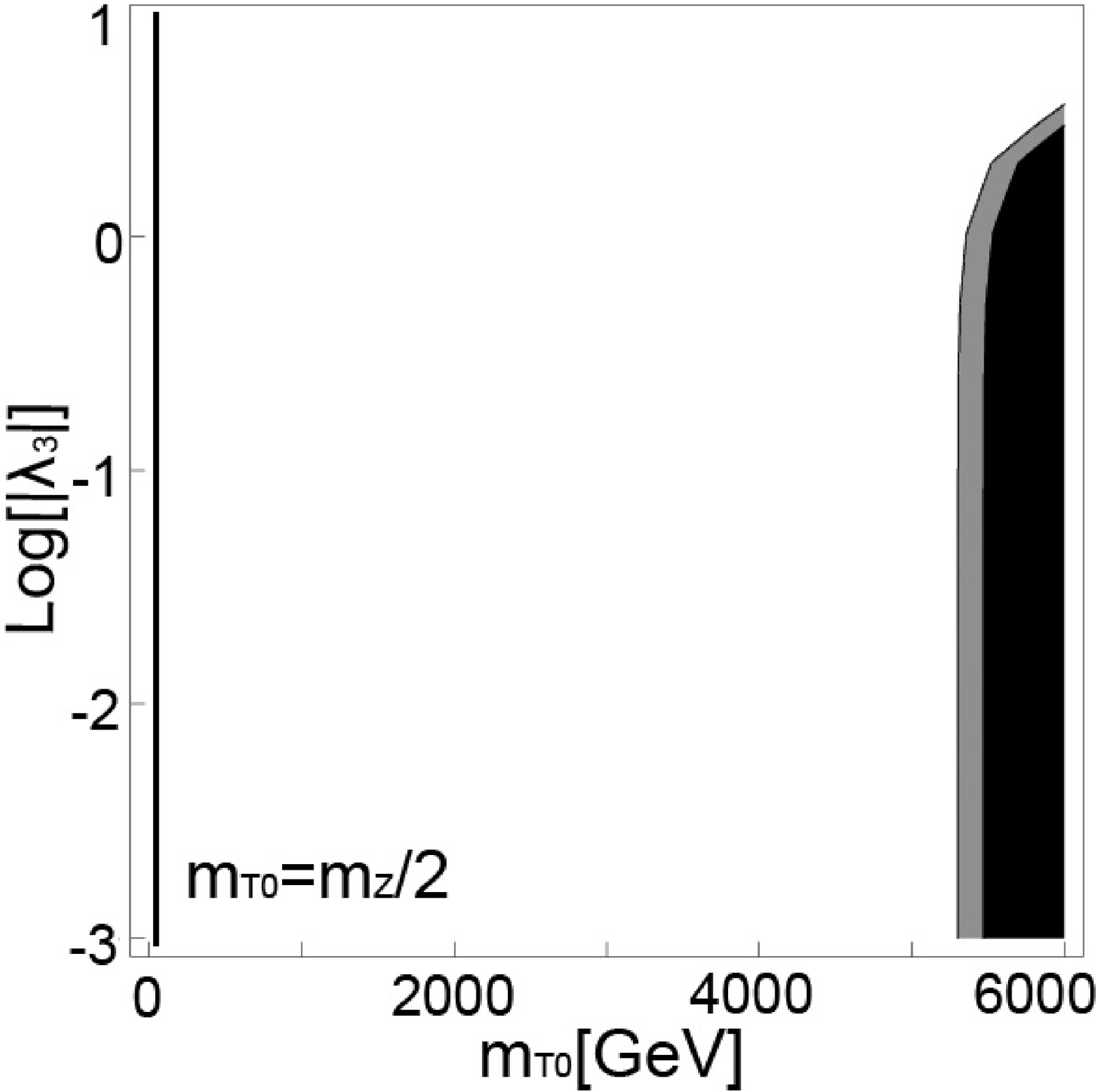}
\caption{Relic abundance for $m_h=120~\GeV$ with the vertical axis of $\mathrm{Log}[\lmd_{3}]$, where black, gray, and white regions show the parameter regions larger than, agrees with, and smaller than the WMAP constraint, respectively.}
\label{fig:relic_abundance2}
\end{center}
\end{figure}
In Figs. \ref{fig:relic_abundance} and \ref{fig:relic_abundance2}, we show the relic abundance of \,$\tz$, where we 
have used \textit{micrOMEGAS\,2.4}\cite{Belanger:2010gh}.
%
For small couplings, i.e. $\lmd_3 \lesssim 1$, the dark matter annihilation is governed by the
weak interaction. So the annihilation cross section does not decrease so much.
\ignore{In this case, the main (co)annihilation modes are $\tz\,T^\pm\to \gamma W^\pm$,
$\tz\,\tz \to W^+W^-$ and $T^+\,T^-\to W^+W^-$.}
On the other hand, in the large coupling region (i.e. $\lmd_3 \gtrsim 1$), the main annihilation modes are
Higgs interactions.
Since the trilinear coupling of $h$ involves only $\lambda_3$,
the cross sections are enhanced by $\lmd_3$.
%
%
From the figure, we find that for $5.4\ \TeV \lesssim m_\tz \lesssim 6\ \TeV$, the relic abundance agrees with 
the WMAP data in Eq.~(\ref{eq:WMAP_relic}).


\begin{figure}[tbp]
\begin{center}
\includegraphics[width=6cm,angle=0,clip]{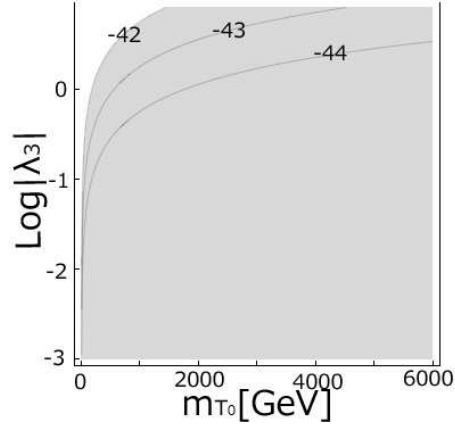}
\caption{Spin-independent scattering cross section between the dark matter $\tz$ and nucleus particles, where numbers on lines represent the cross sections in $cm^{2}$ unit, while the light gray region is allowed by direct searches of dark matter. }
\label{fig:direct_detection}
\end{center}
\end{figure}
The SI cross section of the Y=0 ITM is shown in Fig. \ref{fig:direct_detection}.
From the figure, we can see that in most of the region, the model escapes the constraint from the direct search.
%
In this model, the allowed  processes are the $\tz$-quark (u,d) scatterings at tree level with the small Yukawa coupling,
and $\tz$-gluon scatterings at loop level. 
As a result, the SI cross section is clearly suppressed.

\subsection{Dark Matter in the model of Y=2}

\begin{figure}[t]
\begin{center}
\includegraphics[width=10.5cm,clip]{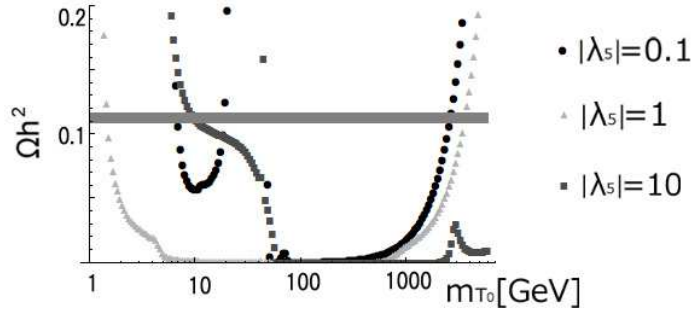}
\caption{Relic abundance in the Y=2 ITM for $m_h=120\GeV$ with the 
vertical axis of $\mathrm{Log}[|\lmd_{5}|]$, where the light gray region is  allowed by WMAP.
}
\label{fig:relic_abundance100}
\end{center}
\end{figure}
\begin{figure}
  \begin{center}
    \begin{tabular}{cc}
      \resizebox{60mm}{!}{\includegraphics{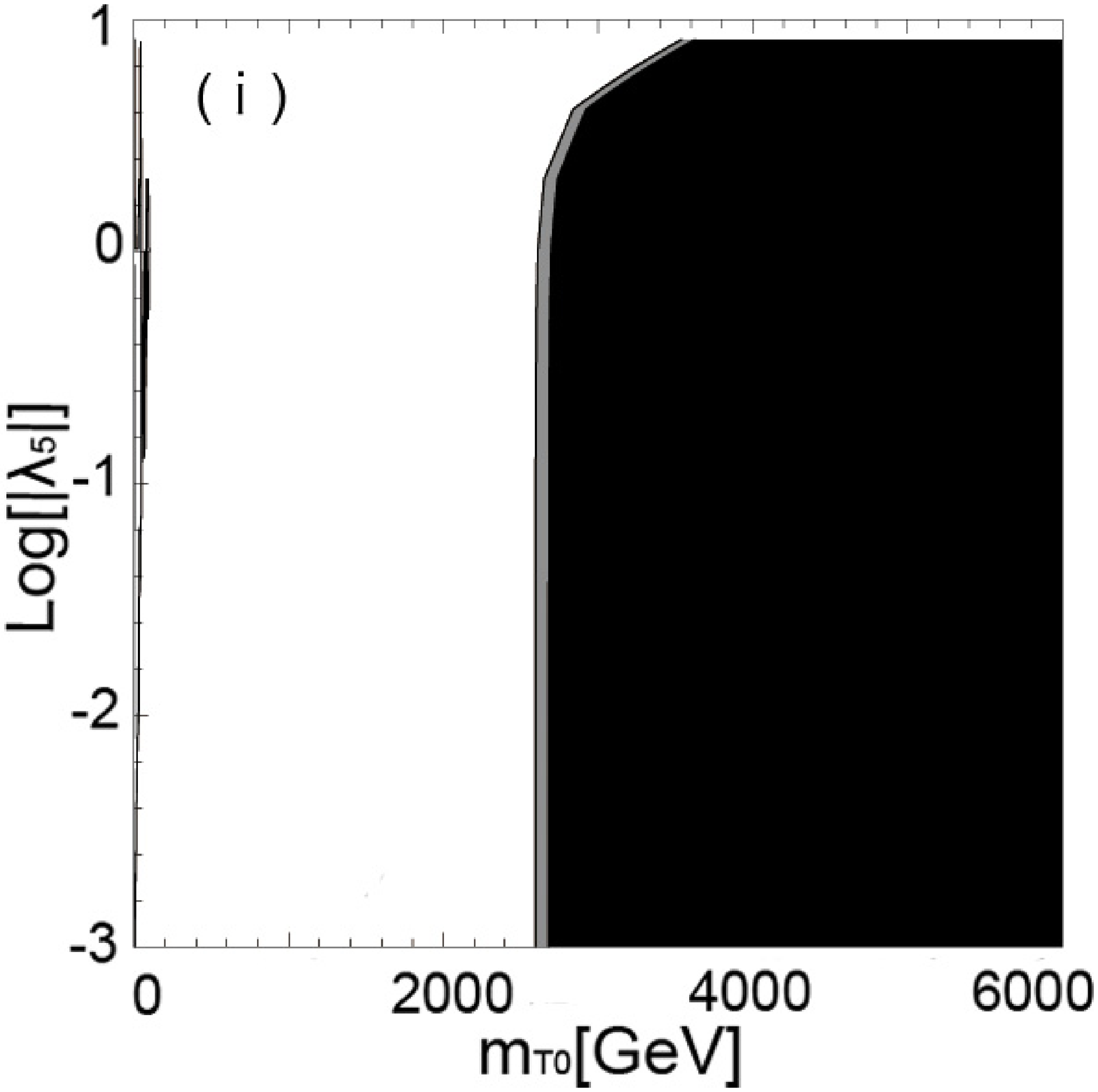}} &
      \resizebox{60mm}{!}{\includegraphics{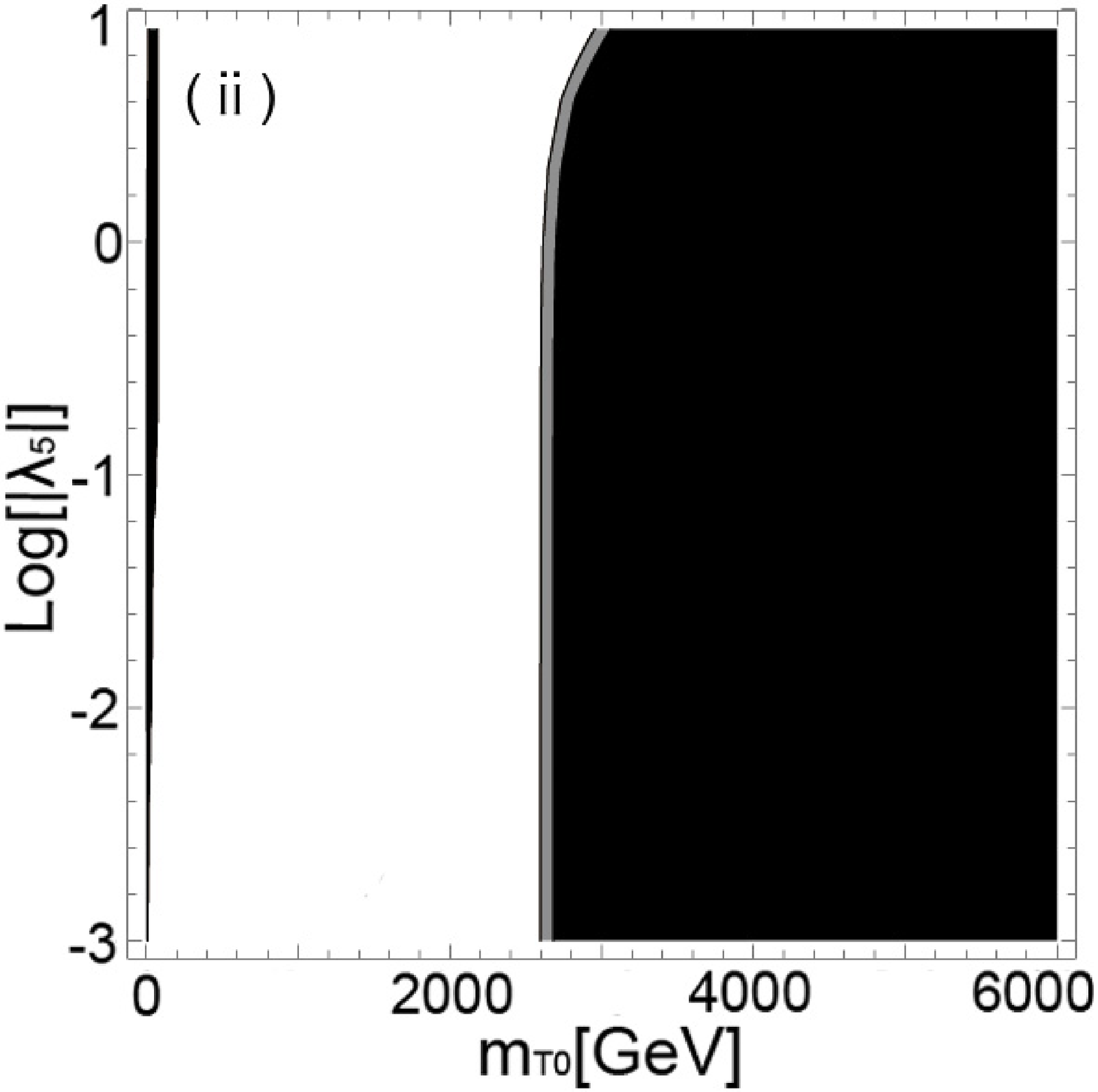}} \\
      \resizebox{60mm}{!}{\includegraphics{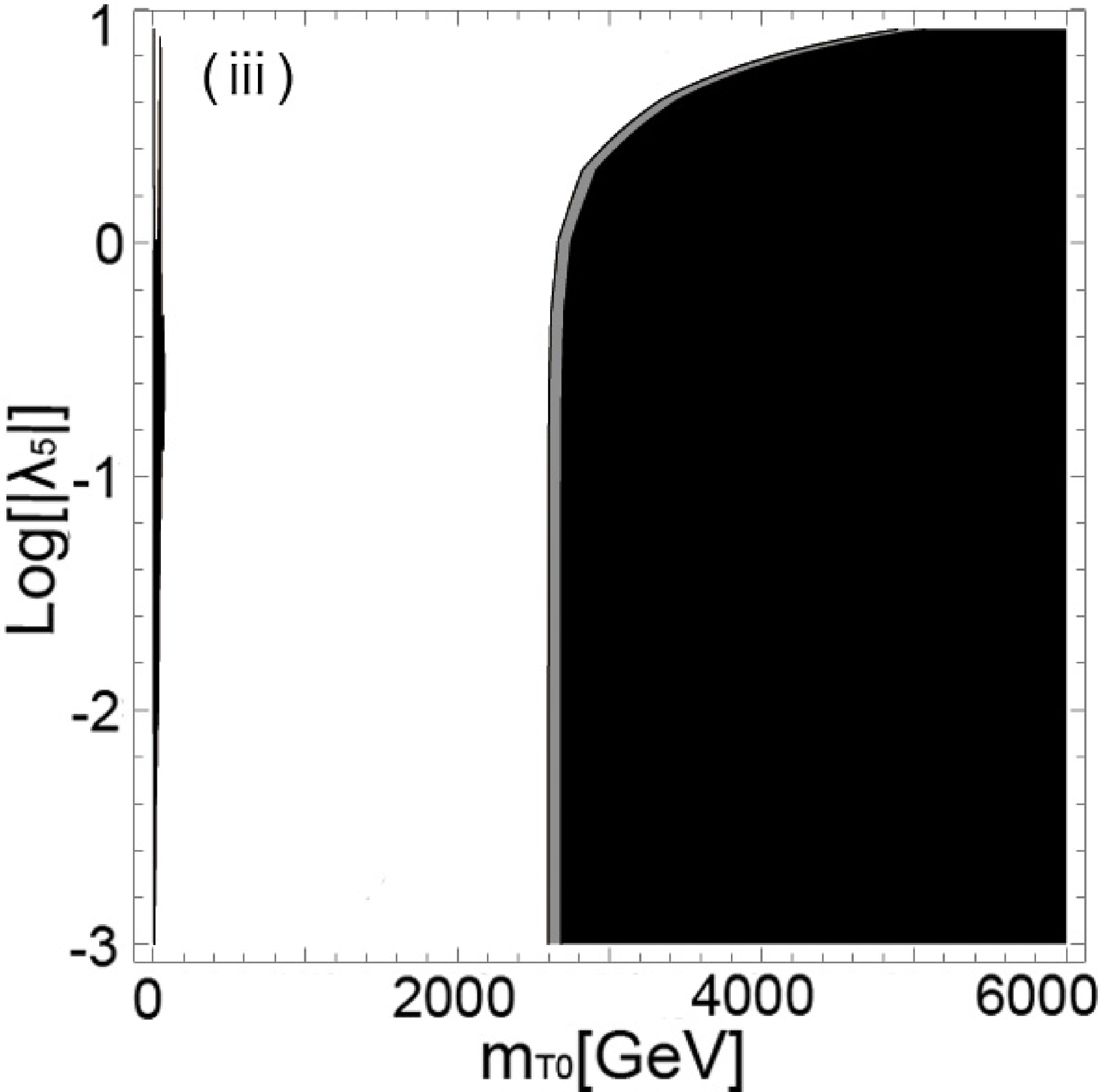}} &
      \resizebox{60mm}{!}{\includegraphics{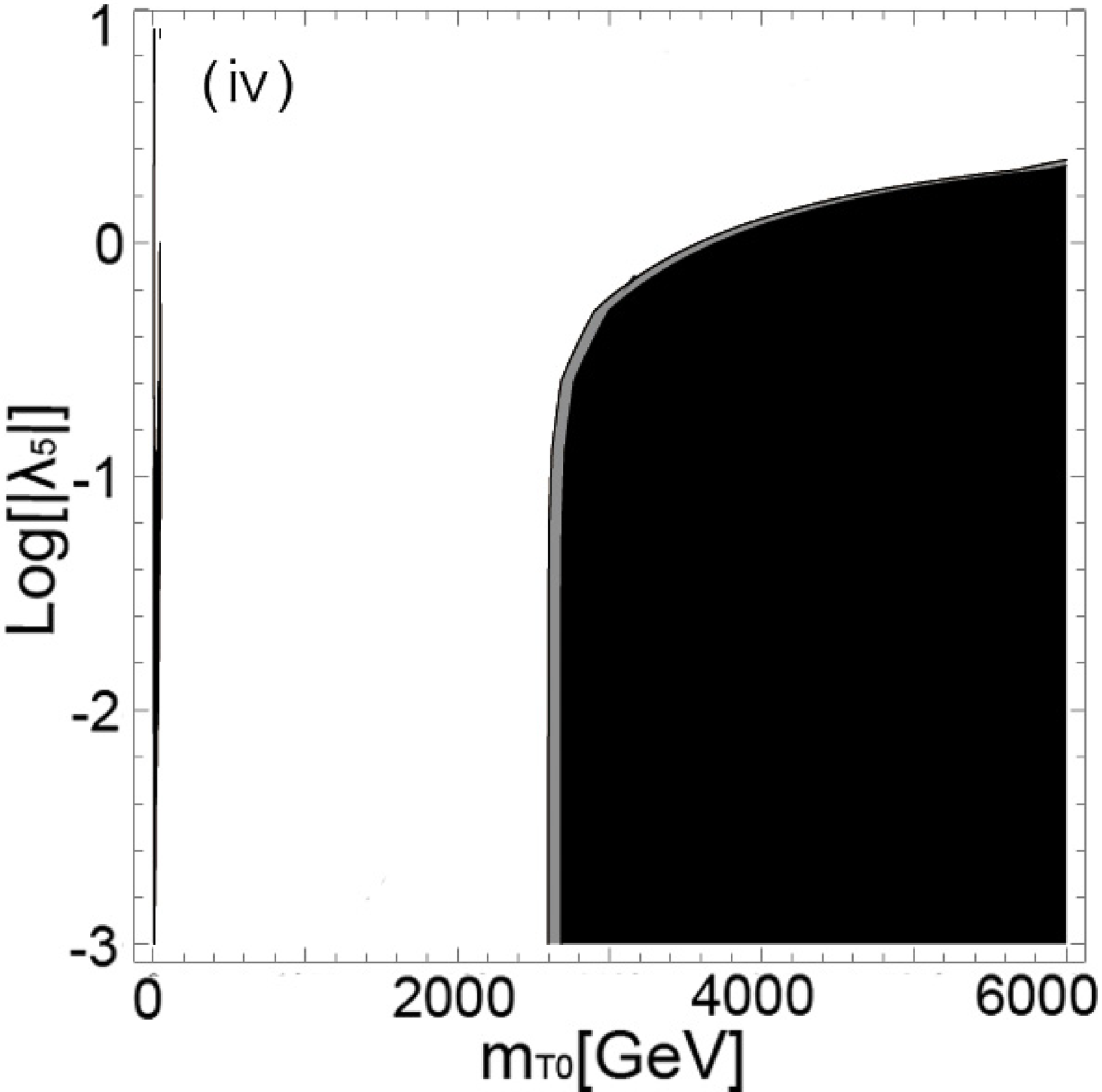}} \\
    \end{tabular}
    \caption{Legend is same as Fig.\ref{fig:relic_abundance2}, but for the Y=2 inert triplet model,
    where (i) $\lmd_4=|\lmd_5|/8$, (ii) $\lmd_4=|\lmd_5|$, (iii) $\lmd_4=2|\lmd_5|$, and (iv) $\lmd_4=8|\lmd_5|$. }
    \label{fig:relic_abundance3}
  \end{center}
\end{figure}

The total relic abundance of $\tzr$ and $\tzi$ is shown
in Fig. \ref{fig:relic_abundance3}.
Note that the masses of $T^\pm$ and $T^{\pm\pm}$ are automatically fixed if $m_{\tzr}$ and $\lmd_5$ are known.
%
Since the degeneracy of triplet masses is lifted at tree level,
coannihilations of the triplet scalars are not so effective compared to that in the  $Y=0$ case.
Thus, the relic abundance gets enhanced.
In the large $|\lmd_5|$ region, as the mass degeneracy of the triplet components is lifted, the coannihilation effect becomes weaker 
 than that in the small $|\lmd_5|$,
which enhances the relic abundance.
However, the annihilation cross section becomes large due to the large couplings of  $\lmd_4$ and $|\lmd_5|$, which suppresses the relic abundance more effective than the coannihilation effect.

 From Fig.~\ref{fig:relic_abundance100}, one can see that in the small mass region,
the relic abundance drastically changes due to the resonance effect 
as well as the opening of new annihilation final states.
In the figure, we have  fixed $\lmd_4=|\lmd_5|/8$. 
%
As $\lmd_4$ approaches to $|\lmd_5|$, the figures become similar to those with small $|\lmd_5|$
since the main interactions enhanced by the large Higgs coupling are proportional to $(\lmd_4+\lmd_5)$.
In the region with $|\lmd_4|/|\lmd_5|\neq 1$, where $\tzr \,\tzr (\tzi \tzi)\to h\,h$  is most effective,
the relic abundance is reduced.
In the case of $|\lmd_4|/|\lmd_5|= 1$, the tri-Higgs couplings proportional to $(\lmd_4+\lmd_5)$ are canceled to $0$.
Thus, the effective couplings of the Higgs bosons are very weak and 
the relic abundance is mainly determined by gauge interactions.

We comment on the direct detection of the Y=2 case.
Unlike those in  Y=0, there are additional scattering processes in the Y=2 model,
which are the $\tz$-quark scatterings through the gauge coupling of $\tz$ to Z.
They have larger cross sections due to  the gauge coupling.
Because of these large cross sections, almost  all region is excluded by the direct detection constraint in Eq.~(\ref{eq:DD_constraint}).
Note that since the scattering process does not depend on the Higgs coupling, this tendency is same even if the ratio of $\lmd_4$ and $|\lmd_5|$ changes.
%

\section{Conclusion}
\label{sec:conclusion}
We have studied dark matter in the two inert triplet models and we have found that
there are allowed regions which agree with the WMAP result in \tev scale for both models.
Explicitly,we have shown that in the Y=0  model, the dark matter mass of the neutral scalar around $m_\tz \sim 5.5\TeV$ is favored by WMAP,
which is also allowed by the direct detection,
while in the Y=2 model, 
$m_\tz\sim 2.8\, \TeV$ is preferred in terms of the relic abundance, but most of the region is excluded by the direct detection constraint 
since the $\tz$-quark scattering cross section mediated by Z enhanced the SI cross section,


\section*{Acknowledgement}
We are grateful to G.~B$\mathrm{\acute{e}}$langer and A.~Pukhov for their kind help for {\it micrOMEGAs}.
The work of T.A. was supported in part by the National
Natural Science Foundation of China under Grant No. 10425522 and No.
10875131. C.Q.G. and K.I.N were partially supported  by the National Science
Council of Taiwan under Grant No. NSC-98-2112-M-007-008-MY3 and the National
Tsing Hua University under the Boost Program No. 97N2309F1.


\begin{thebibliography}{99}
\bibitem{ITM}
 T.~Araki, C.~Q.~Geng and K.~I.~Nagao,
  Phys.\ Rev.\  D {\bf 83}, 075014 (2011)
  [arXiv:1102.4906 [hep-ph]].
 
\bibitem{Komatsu:2010fb}
  E.~Komatsu {\it et al.},
  arXiv:1001.4538 [astro-ph.CO].
 
  \bibitem{Silveira:1985rk}
  V.~Silveira and A.~Zee,
  Phys.\ Lett.\  B {\bf 161}, 136 (1985);
  C.~P.~Burgess, M.~Pospelov and T.~ter Veldhuis,
  Nucl.\ Phys.\  B {\bf 619}, 709 (2001);
  W.~L.~Guo and Y.~L.~Wu,
  JHEP {\bf 1010}, 083 (2010).
\bibitem{Ma:2006km}
  E.~Ma,
  Phys.\ Rev.\  D {\bf 73}, 077301 (2006);
  R.~Barbieri, L.~J.~Hall and V.~S.~Rychkov,
  Phys.\ Rev.\  D {\bf 74}, 015007 (2006);
  L.~Lopez Honorez, E.~Nezri, J.~F.~Oliver and M.~H.~G.~Tytgat,
  JCAP {\bf 0702}, 028 (2007);
  E.~Lundstrom, M.~Gustafsson and J.~Edsjo,
  Phys.\ Rev.\  D {\bf 79}, 035013 (2009);
  P.~Agrawal, E.~M.~Dolle and C.~A.~Krenke,
  Phys.\ Rev.\  D {\bf 79}, 015015 (2009);
  S.~Andreas, M.~H.~G.~Tytgat and Q.~Swillens,
  JCAP {\bf 0904}, 004 (2009);
  E.~Nezri, M.~H.~G.~Tytgat and G.~Vertongen,
  JCAP {\bf 0904}, 014 (2009);
  E.~M.~Dolle and S.~Su,
  Phys.\ Rev.\  D {\bf 80}, 055012 (2009);
  C.~Arina, F.~S.~Ling and M.~H.~G.~Tytgat,
  JCAP {\bf 0910}, 018 (2009);
  E.~Dolle, X.~Miao, S.~Su and B.~Thomas,
  Phys.\ Rev.\  D {\bf 81}, 035003 (2010);
  X.~Miao, S.~Su and B.~Thomas,
  Phys.\ Rev.\  D {\bf 82}, 035009 (2010);
  L.~Lopez-Honorez and C.~E.~Yaguna,
  arXiv:1011.1411 [hep-ph].

\bibitem{Cirelli:2009uv}
  M.~Cirelli and A.~Strumia,
  New J.\ Phys.\  {\bf 11}, 105005 (2009).
  
\bibitem{Hambye:2009pw}
  T.~Hambye, F.~S.~Ling, L.~Lopez Honorez and J.~Rocher,
  JHEP {\bf 0907}, 090 (2009)
  [Erratum-ibid.\  {\bf 1005}, 066 (2010)].

\bibitem{Angle:2007uj}
  J.~Angle {\it et al.}  [XENON Collaboration],
  Phys.\ Rev.\ Lett.\  {\bf 100}, 021303 (2008);
  Z.~Ahmed {\it et al.}  [The CDMS-II Collaboration],
  Science {\bf 327}, 1619 (2010).

\bibitem{oblq}
H. H. Zhang, W. B. Yan and X. S. Li,
{\em Mod. Phys. Lett. A} {\bf 23}, 637 (2008). 

\bibitem{LEP}
LEP Electroweak Working Group, http://lepewwg.web.cern.ch/LEPEWWG/.

\bibitem{tev}
 Tevatron New Phenomena and Higgs Working Group, 
http://tevnphwg.fnal.gov/.


\bibitem{Griest:1990kh}
  K.~Griest and D.~Seckel,
  Phys.\ Rev.\  D {\bf 43}, 3191 (1991);
  S.~Mizuta and M.~Yamaguchi,
  Phys.\ Lett.\  B {\bf 298}, 120 (1993).
  
\bibitem{Belanger:2010gh}
  G.~Belanger, F.~Boudjema, P.~Brun, A.~Pukhov, S.~Rosier-Lees, P.~Salati and A.~Semenov,
  arXiv:1004.1092 [hep-ph].
  
\bibitem{Chun:2009mh}
  E.~J.~Chun,
  JHEP {\bf 0912}, 055 (2009).

\end{thebibliography}
\end{document}